\documentclass{aa}  

\usepackage{graphicx}
\usepackage{txfonts}
\usepackage{amsmath}
\usepackage{multirow}
\usepackage{sidecap}

\usepackage[colorlinks]{hyperref}
\hypersetup{
     colorlinks   = true,
     citecolor    = blue,
     linkcolor    = red,
     urlcolor     = magenta
}
\usepackage{orcidlink}

\titlerunning{Galaxy clusters simulations}
\authorrunning{Leuzzi et al.}
\begin{document}

   \title{Observation-driven simulations of strong lensing galaxy clusters}

   \author{L. Leuzzi\orcidlink{0009-0006-4479-7017}\inst{1}\fnmsep\thanks{\email{laura.leuzzi@inaf.it}}
           \and
           M. Meneghetti\inst{1,6}
           \and 
           A. Adam\inst{3, 4, 5}
           \and 
           L. Moscardini\inst{2, 1, 6}
          \and
           C. Giocoli\inst{1,6}       
           \and 
           P. Bergamini\inst{1, 7}
           \and
           Y. Hezaveh\inst{3, 4, 5, 8, 9, 10}
           \and
           M. Maturi\inst{11, 12}
           \and
           A. Mercurio\inst{13, 14, 15}
           \and
           A. Moretti\inst{16}
           \and
           Andrés A. Plazas Malagón\orcidlink{0000-0002-2598-0514}\inst{17,18}
           \and
           P. Rosati\inst{19, 1}
          }

   \institute{INAF--OAS, Osservatorio di Astrofisica e Scienza dello Spazio di Bologna, via Gobetti 93/3, 40129 Bologna, Italy
         \and
             Dipartimento di Fisica e Astronomia "A. Righi", Alma Mater Studiorum Università di Bologna, Via Gobetti 93/2, 40129 Bologna, Italy
        \and
             Ciela -- Montreal Institute for Astrophysical Data Analysis and Machine Learning, Montréal, Canada
        \and
             Mila -- Quebec Artificial Intelligence Institute, 6666 Rue Saint-Urbain, QC H2S 3H1, Montréal, Canada
        \and
            Department of Physics, Université de Montréal, 1375 Ave.Thérèse-Lavoie-Roux, QC H2V 0B3, Montréal, Canada
        \and
            INFN--Sezione di Bologna, Viale Berti Pichat 6/2, 40127 Bologna, Italy
        \and
           Dipartimento di Fisica, Università degli Studi di Milano, via Celoria 16, I-20133 Milano, Italy
        \and
            Center for Computational Astrophysics, Flatiron Institute, New York, USA 
        \and 
            Perimeter Institute for Theoretical Physics, Waterloo, Canada 
        \and
            Trottier Space Institute, McGill University, Montréal, Canada
        \and
            Zentrum für Astronomie, Universität Heidelberg, Philosophenweg 12, D-69120 Heidelberg, Germany
        \and 
           Institut für Theoretische Physik, Universität Heidelberg, Philosophenweg 16, D-69120 Heidelberg, Germany
        \and
            Università di Salerno, Dipartimento di Fisica “E.R. Caianiello”, Via Giovanni Paolo II 132, 84084 Fisciano (SA), Italy.
        \and
            INAF Osservatorio Astronomico di Capodimonte, Salita Moiariello 16, 80131 Napoli, Italy.
        \and
            INFN–Gruppo Collegato di Salerno–Sezione di Napoli, Dipartimento di Fisica “E.R. Caianiello”, Università di Salerno, Via Giovanni Paolo II, 132, 84084 Fisciano (SA), Italy
        \and
           INAF-Osservatorio Astronomico di Padova, Vicolo Osservatorio 5, 35122 Padova, Italy
        \and
           Kavli Institute for Particle Astrophysics and Cosmology, Stanford University, Stanford, California, United States
        \and
           SLAC National Accelerator Laboratory, Menlo Park, California, United States
        \and
           Dipartimento di Fisica e Scienze della Terra, Università degli Studi di Ferrara, via Saragat 1, 44122 Ferrara, Italy
             }
   \date{Received September 15, 1996; accepted March 16, 1997}

  \abstract
   {Galaxy clusters are the most powerful strong lenses: they greatly magnify the flux of distant and faint sources. Strong lensing also allows for the reconstruction of their mass distribution with $\sim1\%$ level accuracy and enables investigating cosmological parameters. The number of known systems of this type is bound to increase in the next years thanks to wide imaging surveys. Using simulations in this context is crucial to validate the analysis methods before they are applied to real data, and to train machine learning algorithms that can handle large volumes of images. In this work, we present a simulated set of one hundred images of galaxy clusters that we have produced with a novel code for simulating cluster-scale strong lenses. One of the main novelties of our approach, distinguishing it from other existing codes, is the use of empirical relations, derived from state-of-the-art observations, for modelling the characteristics, such as morphology, color, and spatial distribution of the cluster member population. This allows us to reliably reproduce the complexity of real observations. The simulations are partly carried out with the latest version of \texttt{SkyLens}, a code that creates mock observations of strong lensing events in different systems and observational setups. The main improvements we introduce are the use of the message passage interface (MPI) paradigm, a standard for parallel programming that leverages the use of several processors to perform some given task, and the implementation of score-based diffusion models to augment the images of the background sources. Together, they lead to more efficient and realistic image simulations. We also present the validation of the code and simulations by comparing the properties of the mock clusters to those of real ones.  We make the images, deflection maps, convergence maps, and catalogues of cluster members and background sources publicly available to the community at this \href{https://openscience.inaf.it/?page_id=724}{link}.}


   \maketitle

\section{Introduction}\label{sec:intro}
Gravitational lensing occurs when a massive object, the lens, deflects the path followed by the photons emitted by a distant source. In the regime of strong lensing, which happens when the observer, the lens, and the source are closely aligned along the line-of-sight (LoS), several effects caused by the deflection are observed. Firstly, the shape of the emitting background source is generally highly distorted into characteristic ring- or arc-shaped features. Moreover, a background source might also be lensed into multiple images. Their fluxes might be magnified or de-magnified by the lens, so that they will appear brighter or fainter than the unlensed background source intrinsically is.

The type and significance of the events depend on several properties of the systems, including the distances between the sources and the lenses, and some of the characteristics of the lens, such as the mass distribution. Both galaxies and galaxy clusters can act as lenses, but in this work, we will focus on galaxy clusters. They are the most powerful known lenses because they are the largest gravitationally-bound structures in the universe \citep[see][for a review on lensing by clusters]{umetsu_2020, natarajan_2024}. 

Studying cluster-scale strong lenses has proven very useful for several applications. The magnification power of the clusters enables the observation of very faint sources, such as galaxies at the epoch of reionisation \citep[e.g.,][]{livermore_2017, relics_2019, pascale_2022} and candidate population III stellar systems \citep[e.g.,][]{vanzella_2020, vanzella_2023}. Moreover, modelling lensing features in the cluster cores is a well-established method for investigating the radial mass distribution and for measuring the total mass enclosed within the cluster’s Einstein ring \citep[e.g.,][]{pignataro_2021, caminha_2022a, furtak_2023}. By combining strong and weak lensing measurements, which trace the mass distribution in the outskirts of the clusters, it is also possible to reconstruct the mass radial profile up to larger radii \citep[e.g.,][]{bradac_2006, merten_2009, okabe_2016, umetsu_2016, schrabback_2021, murray_2022, niemiec_2023, diego_2025, cha_2025}. The investigation of strong lensing by galaxy clusters is also relevant in cosmology. The properties and the counts of the arcs observed in the clusters are sensitive to various cosmological parameters, such as $\Omega_M$, $\sigma_8$, $\Omega_\Lambda$ \citep[e.g.,][]{bartelmann_1998, meneghetti_2013, boldrin_2016}. Moreover, modelling the clusters' mass distribution has been used to estimate the dark energy equation of state \citep[e.g.,][]{sereno_2002, jullo_2010, magana_2018, caminha_2022b}.

The individual galaxies within the cluster cause perturbations in the cluster's large-scale potential and may act as strong lenses if aligned with background sources. For this reason, the appearance of strong lenses on smaller scales provides additional information on the distribution of matter in the central areas of the cluster \citep[e.g.,][]{bergamini_2019}. Furthermore, the system of multiple images of a distant supernova produced by an early-type galaxy in a cluster has been used by \cite{kelly_2015} to further characterise the system. The time delay in the appearance of the multiple images has also been used to constrain the Hubble constant $H_0$ \citep[e.g.,][]{soucail_2004, gilmore_2009, vegaferrero_2018, pascale_2025, liu_2025}.

The number of observed galaxy clusters available for modelling in the next few years will significantly increase. For example, the wide survey that will be carried out by the Euclid telescope \citep{Scaramella-EP1, euclid_overview} during its six-year mission is expected to deliver a catalogue of $> 10^5$ clusters up to $z\sim2$ \citep{sartoris_2016, Adam-EP3}, and thousands of them are expected to show giant arcs \citep{boldrin_2012, boldrin_2016}. Furthermore, the Legacy Survey of Space and Time conducted with the Vera Rubin Observatory is also expected to observe several thousand galaxy clusters, many overlapping with those observed by Euclid, and about 1000 of them \citep{lsst_sciencebook, ivezic_2019, shajib_2025} will show multiple image systems.

In this context, where the number of strong-lensing clusters is expected to increase significantly, and the quality of the observations will improve as well, simulations play a crucial role for several reasons (see \citealt{plazas_2020} for an overview). One of them is the characterization of the systematic uncertainties and the assessment of the accuracy of modelling techniques that will be applied to observational data. For example, \cite{meneghetti_2017} have tested the effectiveness of several modelling methods on simulated clusters, where the underlying mass distribution is known, thus providing a controlled evaluation of their expected performance when applied to real observational data. Moreover, the comparison between the observations and the results of simulations is a powerful tool to validate cosmological models. In this context, for example, \cite{meneghetti_2020} have found a discrepancy between the observed number of galaxy-scale strong lenses in the core of clusters and that predicted by the $\Lambda$-cold dark matter ($\Lambda$CDM) paradigm, that has been studied in subsequent works \citep[see][]{ragagnin_2022, meneghetti_2022, meneghetti_2023}, but is still not solved. Lastly, the optimal exploitation of future measurements of cluster lensing will require a shift in the methods used for modelling the mass distribution of the clusters.  Given the enormous volume of data imaging surveys produce, these techniques will increasingly rely on automation and Machine Learning (ML) \citep[e.g.,][]{avestruz_2019}. In this scenario, the simulation of highly realistic images will also be necessary for training the algorithms for detecting and modelling strong lenses \citep[e.g.,][]{ntampaka_2019}. 

In this work, we present an improved code to simulate the images of galaxy clusters, its validation and a simulated set of images we release. We simulate the lensing events with \texttt{SkyLens} \citep{meneghetti_2007, meneghetti_2008, meneghetti_2010, skylens_plazas}. This well-established code can handle various observational configurations and simulate the effects of different physical systems, namely galaxies and galaxy clusters. We use empirical relations to determine the physical properties of the cluster members, and we link these properties to those of the main dark matter halo of the cluster. The simulation of strong lensing by galaxy clusters is more complicated than that of galaxy-lenses, because of the higher internal complexity of these objects. Thus, from a numerical point of view, one of the key challenges is to have a fast code that can generate a large number of images in a reasonable amount of time, while keeping a good level of realism. For this reason, we introduce the Message Passage Interface (MPI) formalism in our code. This accelerates the computation of the deflection angle maps and the image generation. In particular, we have produced a set of one hundred images that are available to the community at this \href{https://openscience.inaf.it/?page_id=724}{link}, along with a catalogue of the main properties of the cluster members and background sources, compiled as explained in the \texttt{readme} file, and the deflection and convergence maps of the systems.

This paper is organised as follows: in Sect. \ref{sec:grav_lensing} we explain how lensing affects the images of background sources in the strong lensing regime; in Sect. \ref{sec:skylens} we introduce \texttt{SkyLens} and the MPI formalism; in Sect. \ref{sec:our_dataset} we present our choices for the simulated objects; in Sect. \ref{sec:frontierfields} we validate the simulation procedure; in Sect. \ref{sec:discussion} we discuss potential applications of the simulations, concluding in Sect. \ref{sec:conclusions}. Throughout this work, we assume a flat $\Lambda$CDM cosmological model, with $\Omega_{m,0} = 0.3$ and $H_0 = 70$~ km/s/Mpc. All computational times reported in this work are based on tests performed on a multi-core machine with an Intel(R) Xeon(R) Silver 4214 CPU at 2.20GHz, using 16 cores. The number of cores can be adjusted according to the available computational resources.

\section{Gravitational lensing}\label{sec:grav_lensing}
In the strong lensing regime, the shapes of the background sources are distorted, and their observed positions on the sky are shifted. These effects are measured by the deflection angle $\vec \alpha$, which is related to the difference between the original position of the source, $\vec \beta$, and the position of the same source on the image plane, $\vec \theta$. The mapping between the two is expressed through the lens equation
\begin{equation}
    \vec \beta = \vec \theta - \frac{D_{LS}}{D_S}\vec \alpha (\vec \theta),
    \label{eq:lens_equation}
\end{equation}
where $D_S$ is the angular diameter distance between the source and the observer and $D_{LS}$ is the angular diameter distance between the lens and the source. The computation of the deflection field relies on the thin lens approximation \citep{schneider_1992}, where the lens is described as a surface mass density lying on the lens plane. This is well-justified in virtually all cases in astrophysics because the physical size of the mass distribution of the lens along the LoS is small compared to the distance between the observer, the lens, and the source. 

Furthermore, Eq. \ref{eq:lens_equation} accounts for the deflection of one individual lens. A more accurate approximation is the multiple-lens configuration, in which big volumes of matter are projected onto a series of consecutive planes perpendicular to the LoS. This is still an approximation, but it takes into account the presence of lensing matter at different distances from the observer and sources. In this case, the lens equation is written as 
\begin{equation}
\vec \beta = \vec \theta_1 - \sum_{i=1}^{N_S} \vec \alpha_i (\vec \theta_i),
    \label{eq:lens_equation_multiplane}
\end{equation}
\noindent and
\begin{equation}
    \vec \theta_i = \vec \theta_1 - \sum_{j=1}^{i-1} \frac{D_{ji}}{D_i}\frac{D_S}{D_{jS}}\vec \alpha_j (\vec \theta_j)
    \label{eq:theta_i}
\end{equation}
the index $i$ accounts for the planes in which the mass is projected along the LoS, and $\theta_i$  is the angular position at which the light ray intersects the $i$-th lens plane. The effective deflection is given by the sum of the contributions of the matter on all planes.

Aside from their positions being shifted, the background sources are also distorted and magnified. The distortion of each infinitesimal element of the images is described by two local quantities on the image plane \citep{meneghetti_libro}. The convergence $\kappa$ is responsible for an isotropic distortion of the image, which is rescaled in all directions by the same factor. Given the deflection, the convergence is computed as
\begin{equation}
    \kappa(\vec \theta) = \frac{1}{2} \left(\frac{\partial \alpha_1}{\partial \theta_1} + \frac{\partial \alpha_2}{\partial \theta_2}\right).
    \label{eq:convergence}
\end{equation}
Here, the subscripts 1,2 refer to the two different spatial directions of the system.

The second quantity that describes the distortion is the shear $\vec \gamma$, which stretches the intrinsic shape of the source along one privileged direction and has two components:
\begin{equation}
    \gamma_1(\vec \theta) = \frac{1}{2} \left [ \frac{\partial \alpha_1}{\partial \theta_1} - \frac{\partial \alpha_2}{\partial \theta_2}\right ],
    \label{eq:gamma_1}
\end{equation}
\begin{equation}
    \gamma_2 (\vec \theta) = \frac{\partial \alpha_2}{\partial \theta_1} =\frac{\partial \alpha_1}{\partial \theta_2}.
    \label{eq:gamma_2}
\end{equation}
Finally, the deflection of the photons emitted by the background source induces a change in the solid angle covered by the background source, making the observed flux either magnified or demagnified compared to the unlensed flux. The magnification $\mu$ is given by 
\begin{equation}
    \mu = \frac{1}{(1-\kappa)^2 - |\vec \gamma|^2}.
    \label{eq:magnification}
\end{equation}
We refer the reader to \cite{narayan_1996, meneghetti_libro} for more detailed introductions to the strong gravitational lensing phenomenology. 

When simulating a strong lensing event, the greatest computational cost comes from the computation of the deflection field. This is done with ray-tracing methods. Light rays are shot from the position of the observer to the position of the source, and are used to compute the deflection angles, using either the thin lens or the multiplane approximations. Some works \citep[e.g.,][]{killedar_2012, barreira_2016, breton_2022} have presented 3D ray-tracing methods that yield more precise results but are also very computationally expensive. In the case of our simulations, \texttt{SkyLens} implements the multiplane-lens approach, but we adopt the thin-lens approximation, as it was done in other works in the literature \citep[e.g.;][]{pics_simulations}. 

The light rays are shot on a grid that covers the field-of-view (FOV) of the simulation with a given spatial resolution, which depends on the observational setting of the simulation. \cite{metcalf_2014, petkova_2014} have used an adaptive mesh refinement algorithm in the areas where more rays are needed to capture the lensing effects, given the system's properties (i.e., size, location, surface brightness distribution). In the case of galaxy clusters, the total deflection at each point is given by the sum of the deflection caused by the potential of the cluster and the smaller, yet important, contributions of the galaxies within the cluster. Their main effect is to perturb the potential of the cluster's halo, but they also cause an additional lensing signal. 

\section{\texttt{SkyLens}}\label{sec:skylens}
\texttt{SkyLens} is a well-established code that can simulate strong lensing events in various physical configurations and observational settings. It was first introduced in \cite{meneghetti_2008}, who used the simulations to study arc statistics from galaxy clusters. \cite{meneghetti_2010} later applied it to investigate the bias in estimating the masses of clusters derived from X-ray and lensing measurements. \cite{skylens_plazas} presented the latest version of the code, and we refer to this work for a detailed description of the simulation pipeline. Building on this framework, we have implemented specific modifications aimed at improving the efficiency and the computational cost of the simulation process. While the core algorithm remains mostly unchanged, our contributions enable the code’s applicability to large-scale simulations. In this Section, we summarise the main steps of the algorithm and discuss the improvements introduced in our version.

\texttt{SkyLens} requires three input files for the simulation: the deflection angle maps, the catalogue of the cluster members, and the catalogue of background sources. There are several possibilities to generate these inputs: in Sect. \ref{sec:our_dataset}, we present the procedure we follow to create them for the simulations we release along with this paper, using the python-based package \texttt{pyLensLib} \citep{meneghetti_libro}, while we focus on the general pipeline here. 

\texttt{SkyLens} performs the following steps to simulate the lensing events: 

\begin{enumerate}
    \item It reads the properties of the cluster members and background galaxies from the input catalogues, and it generates the population of galaxies in the FOV of the image accordingly. The reader can refer to Sect. \ref{sec:clu_memb}, Sect. \ref{sec:bkg_sources} and Table \ref{tab:clumemb_props} for the full list of these properties and their characterization.
    \item It generates the grid of points on which the images of the galaxies are simulated. In particular, it computes the final position of the images by taking into account the deflection of the light of the sources at their initial position, given by the deflection angle maps.
    \item The input deflection angle maps are rescaled during the simulation to account for the specific redshift of the individual sources in the background field. In particular, the code reads the sources from the catalogue provided as input and divides them into 100 redshift bins in the range $z = 0$ to $z = 12$, whose sizes are defined so that their centres are equally spaced in lensing distance, $D_{LS}D_L/D_S$: to each redshift bin corresponds a source plane. Given that the change in lensing distance is small for small changes in redshift, especially at higher redshifts, this binning is a reasonable trade-off between accuracy and speed. The distortions are computed for the background sources grouped in the source planes defined by the redshift bins.  This process is the same as explained in Sect. 2.5 of \cite{skylens_plazas} and saves computational time compared to considering each source at its redshift.
    \item At every point of the grid, the code evaluates the flux of the background sources and cluster members in a given photometric band, accounting for their spectral energy distributions and the chosen instrumental filter, thus introducing a dependency on the wavelength of the simulation. This procedure is the same as that presented in \cite{skylens_plazas}. Their contributions are summed to obtain a full image of the cluster and the lensed features.
    
    \item Finally, the code convolves the image with the instrumental point spread function (PSF), which is provided as an additional input file containing the effective PSF model for each photometric band of the instrument. This model should also account for atmospheric effects in the case of ground-based simulations.
\end{enumerate}

\begin{figure*}
    \centering
    \includegraphics[width = 0.9\textwidth]{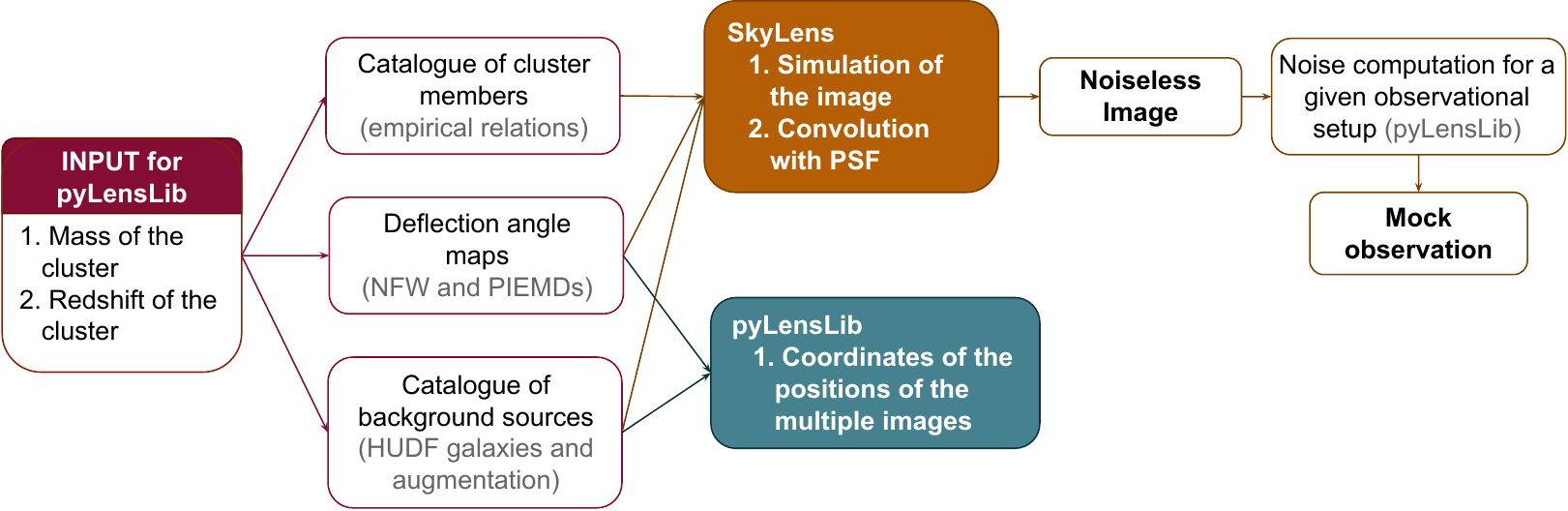}
    \caption{In this flowchart, we outline the pipeline of our simulation. The input of the simulation is the mass and the redshift of the cluster. Depending on them, the code generates the deflection angle maps, the catalogue of cluster members, and background sources. These are used to simulate the image of the cluster in a given filter.}
    \label{fig:flowchart_pipeline}
\end{figure*}

The output of the code is the noiseless image of the cluster and lensed background, which is saved both in its original form and after convolution with the instrument’s PSF. Depending on the observational conditions we want to simulate, we can add noise to the images in a subsequent step implemented with \texttt{pyLensLib} \citep{meneghetti_libro}. In Fig. \ref{fig:flowchart_pipeline}, we show a schematic overview of the pipeline of the simulation we have just presented. 

The simulation of lensing events on cluster scales is computationally expensive. For a rough estimate of the computation time, we can take into account a FOV of $\sim 180\arcsec - 200\arcsec$, consistent with the FOV of the Hubble Space Telescope (HST), the Euclid telescope and the James Webb Space Telescope \citep[JWST;][]{gardner_2006}, since most strong-lensing analyses focus on cluster regions of this size. The chosen FOV sets the number of pixels and rays to be simulated, directly influencing the computational cost. With a spatial resolution of $\sim0.05\arcsec-0.1\arcsec$, which is characteristic of space-based observations, the images we simulate will have millions of pixels. 
Simulating thousands of images with the original version of \texttt{SkyLens} would not be feasible in a reasonable amount of time because ray-tracing the deflections at different redshifts is computationally expensive. For this reason, we also parallelise this procedure using the MPI paradigm. In this paradigm, we define one main process and several child processes that simultaneously perform the same operations on different parts of the data. This reduces the computation time of the calculations proportionately to the number of processors on a given hardware. In our case, we divide the full grid of pixels into 16 sections, one per core in our tests. This number can be adjusted depending on the processors available on the machine on which the code is run and on the size of the images. Each of these sections is paired with a different computational core, and they all perform the same operations. Namely, the fluxes in the pixels of different slices are computed separately by the processes, and the results are gathered by the main process at the end of the simulation to get the full image. After rewriting it in this paradigm, the code needs approximately 5s to generate the image of one cluster in one observational band. This corresponds to a speed-up factor of $\sim 12$, given that the previous version of the code required $\sim 60$s to simulate an equivalent image.

\section{Simulations}\label{sec:our_dataset}
Along with this paper, we make publicly available the simulated images of one hundred clusters, together with the corresponding catalogues of cluster members and of the lensed background galaxies appearing in the FOV. In the following subsections, we describe the generation of the properties of the cluster members and background galaxies, of the lensing properties of the clusters, and the general characteristics of the images.

\subsection{General properties}\label{sec:clusters_props}
We simulate the clusters in the redshift range 0.3 -- 0.6. In particular, we place 30 of them at $z = 0.3$, 30 at $z = 0.4$, 20 at $z = 0.5$ and 20 at $z = 0.6$. We focus on this redshift range because it corresponds to the peak of observing strong lensing events. Their masses are uniformly distributed in the range $M_{200} \sim 5\times10^{14} - 2\times10^{15}M_{\odot}$, where $M_{200}$ is the mass of the cluster within the radius at which its mean density is 200 times the critical density of the Universe at the cluster's redshift, namely $R_{200}$. This mass range covers a wide variety of systems, ranging from galaxy groups to massive galaxy clusters (as in, for example, \citealt{fedeli_2008}).

\begin{figure*}
    \sidecaption
    \includegraphics[width=12cm]{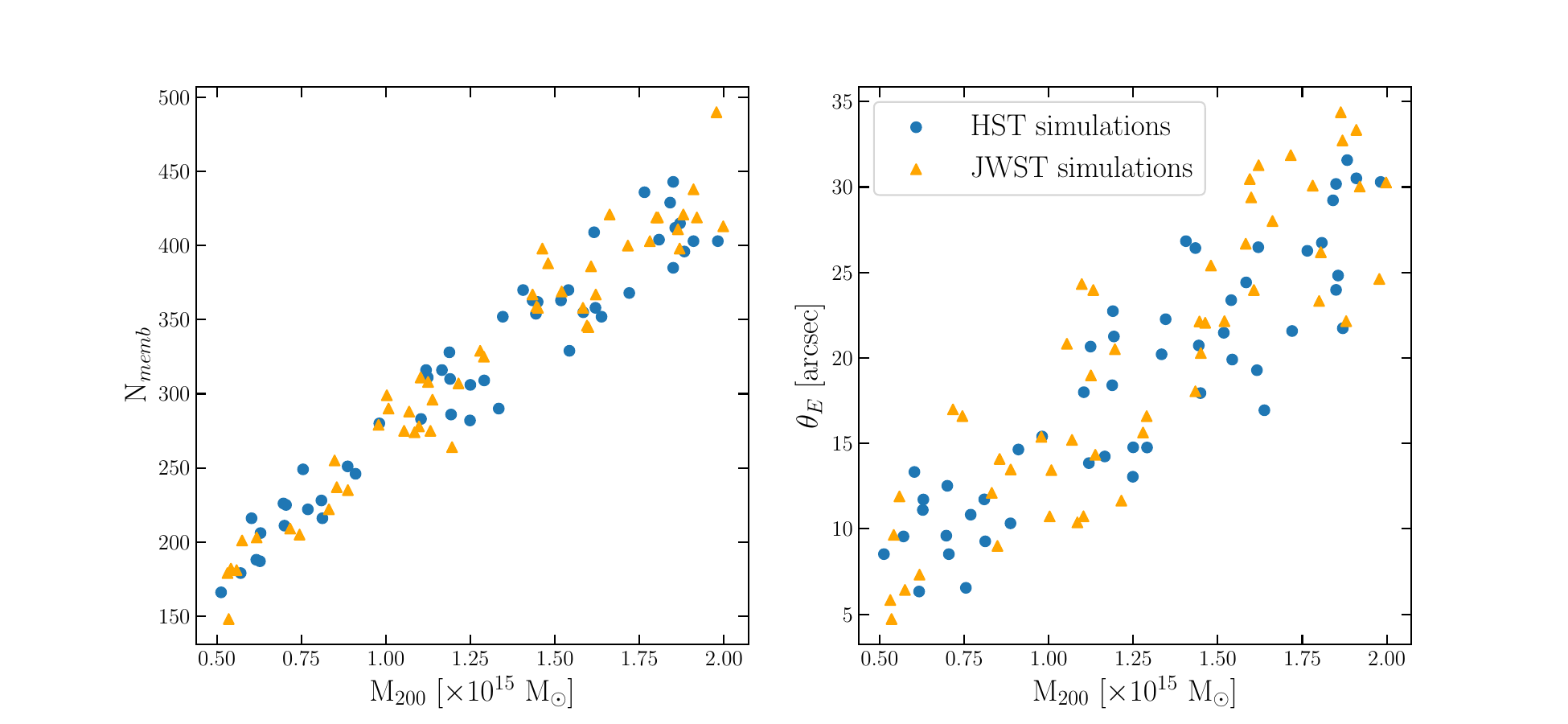}
    \caption{In the left panel, we show the relation between the distribution of the number of cluster members and the mass of the clusters: more massive clusters have a higher number of cluster members. In the right panel, we show the distribution of the main Einstein radius of the simulated clusters as a function of their mass.}
    \label{fig:clusters_plot}
\end{figure*}

We generate two sets of 50 simulated clusters, one assuming the observational characteristics of the JWST and the other those of the HST. In the case of the HST, we simulate the images in different bands: F105W, F125W, F140W, F160W of the Wide Field Camera 3 (WFC3) instrument and F435W, F475W, F606W, F625W, F775W, F814W and F850LP of the Advanced Camera for Survey (ACS) instrument. For both HST/WFC3-like and HST/ACS-like simulations, the FOV is $200\arcsec\times200\arcsec$, with a spatial resolution of $0.05\arcsec$, which results in $4000\times4000$ pixels wide images. The parameters (i.e, sky background level and exposure time) we use to add noise to the images are similar to those of the Cluster Lensing And Supernova survey with Hubble \citep[CLASH;][]{postman_2012}. We summarise them in Table \ref{tab:data_noise}. In the case of the JWST, we simulate the images in the F277W, F356W, F410M, F444W, F090W, F115W and F200W of the Near Infrared Camera (NIRCam): the exposure times and background levels adopted for the noise computation are also in Table \ref{tab:data_noise}. For these images, we choose a FOV of $210\arcsec\times210\arcsec$, which is representative of the regions where strong lensing features are observed and, given the spatial resolution of $0.07\arcsec$, corresponds to $3000\times3000$ pixel images. In both setups, the sky background is simulated as a random, non-correlated Gaussian field that corresponds to the average background level given in Table \ref{tab:data_noise}, and is added to the photon noise of the sources. 

\begin{table}[]
    \centering
    \caption{Summary of the observational setup of the simulated images.}
    \begin{tabular}{c|c|c|c}
        INSTRUMENT & BAND & TEXP (s) & BKG\_MAG \\
         \hline
         HST/ACS & F435W & 1984 & 25.73  \\
         HST/ACS & F475W & 1994 &  25.23 \\
         HST/ACS & F606W & 1975 &  24.82 \\
         HST/ACS & F775W & 2022 &  24.55 \\       
         HST/ACS & F814W & 4103 &  25.43 \\         
         HST/ACS & F850LP & 1984 &  25.22 \\         
         HST/WFC3 & F105W & 2645 & 24.18 \\
         HST/WFC3 & F125W & 2425 & 24.09 \\
         HST/WFC3 & F140W & 2342 & 23.97 \\
         HST/WFC3 & F160W & 4920 & 24.52  \\
         JWST/NIRCam & F277W & 1800 & 22.61 \\
         JWST/NIRCam & F356W & 1800 & 22.61 \\
         JWST/NIRCam & F410M & 1800 & 22.61 \\
         JWST/NIRCam & F444W & 1800 & 21.78 \\
         JWST/NIRCam & F090W & 1800 & 21.82 \\ 
         JWST/NIRCam & F115W & 1800  & 21.85 \\
         JSWT/NIRCam & F200W & 1800 & 22.07
         \end{tabular}
    \tablefoot{The first and second columns describe the instrument and band in which the simulations are done, respectively; the third column is the exposure time, and the fourth column gives the average value of background level in magnitude.}
    \label{tab:data_noise}
\end{table}

In the left panel of Fig. \ref{fig:clusters_plot}, we show the distribution of the
number of cluster members (see \ref{sec:clu_memb} for more details on the generation of this number), while in the right panel of the same Figure we show the Einstein radii of the clusters as a function of their mass. For a spherically symmetrical lens, the Einstein radius is given by
\begin{equation}
    \theta_E = \sqrt{\frac{4GM(\theta_E)}{c^2}\frac{D_{LS}}{D_LD_S}},
    \label{eq:ein_radius}
\end{equation}
and it depends on the mass enclosed within the central region of the lens. It corresponds to the tangential critical line of the lens, and it is a region of very high magnification.

\begin{figure}
    \centering
    \includegraphics[width=0.99\linewidth]{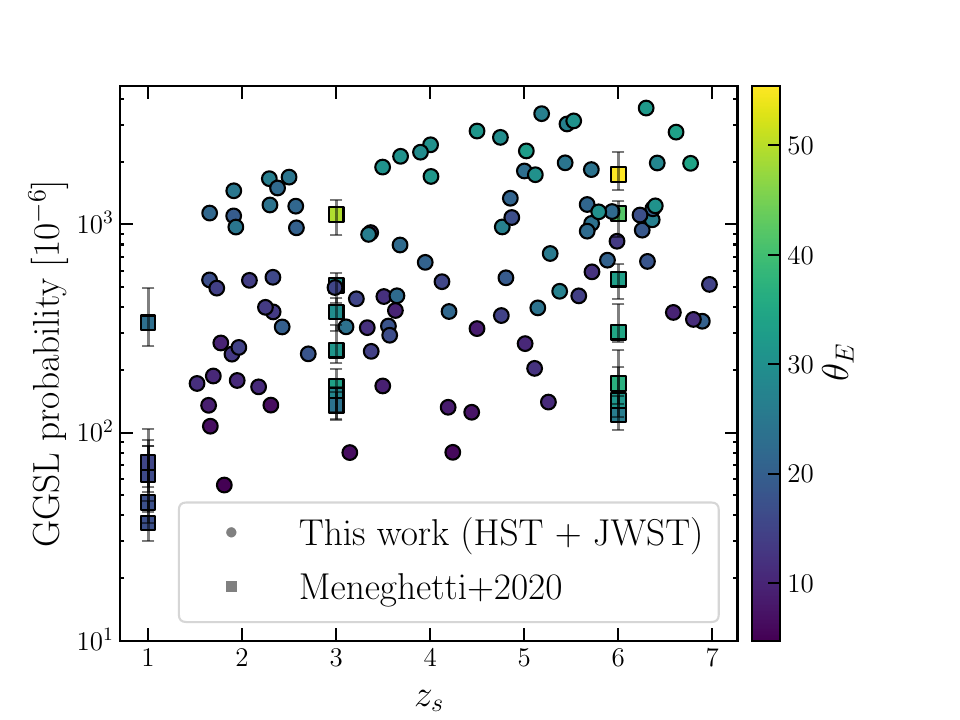}
    \caption{Comparison of the galaxy-scale strong lensing probability (on the y axis) of the simulated clusters released with this paper with that of the clusters studied in \cite{meneghetti_2020}. The probability is shown as a function of the redshift of background sources $z_S$, and the clusters are color-coded for the size of their main critical line expressed as the Einstein radius in arcseconds, $\theta_E$.}
    \label{fig:p_ggsl}
\end{figure}

In Fig. \ref{fig:p_ggsl}, we compare the probability of having galaxy-scale strong lenses in the simulated clusters with that of the observations presented in \cite{meneghetti_2020}. We find good agreement between the values predicted by our simulations and the observations.

\subsection{Mass modeling}\label{sec:mass_model}
The number and type of lensing features that appear in the images depend on the concentration and radial distribution of the mass of the lens: using different mass models yields different results. For our simulations, we model the mass profile of the clusters as the combination of a main dark matter (DM) halo, which represents the cluster's potential, and an ensemble of subhaloes that represent the cluster members. We model their profiles and compute the corresponding deflections with \texttt{pyLensLib}. 
This follows the general ideas developed in the \texttt{MOKA} code \citep{giocoli12a,giocoli16b}.

We parametrise the main halo as an elliptical mass distribution with a Navarro-Frenk-White (NFW) density profile \citep{nfw_profile},
\begin{equation}
    \rho(r) = \frac{\rho_s}{(r/r_s)( 1 + r/r_s)^2},
    \label{eq:nfw_profile}
\end{equation}
where $\rho_s$ is a characteristic density and $r_s$ is the scale radius. In particular, $\rho_s$ is related to $M_{200}$ \citep{limousin_2005}. Thus, we can describe the main halo with $M_{200}$ and $r_s$, in addition to the position angle PA, the axis ratio $q$, and the redshift $ z_l$. 

Moreover, we model the cluster members as pseudo-isothermal elliptical mass distributions (PIEMDs), that is a common choice to model the mass profile of galaxy-scale haloes in the literature \citep[e.g.,][]{magana_2018, bergamini_2019, acebron_2020, pignataro_2021, galan_2025}, because it accounts for the ellipticity of the galaxy-lenses and, in contrast to the singular isothermal sphere model, has a finite total mass distribution. The density profile of this model is given by
\begin{equation}
    \rho(r) = \frac{\rho_0}{(1 + r^2/r_{core}^2)(1 + r^2/r_{cut}^2)}.
    \label{eq:piemd}
\end{equation}
The parameters that define the PIEMDs are the central density $\rho_0$, that can be expressed in terms of the central velocity dispersion $\sigma_0$ \citep{limousin_2005}, the core radius $r_{core}$ and the cut radius $r_{cut}$. Moreover, each subhalo has its own position angle (PA) and axis ratio $q$. In particular, we derive $\sigma_0$ and $r_{cut}$ from their photometric properties, as given in eq. \ref{eq:sigma_lum} and eq. \ref{eq:radius_sigma}, that relate these properties with the luminosity of the galaxies (see Sect. \ref{sec:clu_memb} for more details on the generation of the cluster members' properties). 

Once the mass distribution has been defined, we compute the deflection field normalised for a background source at $ z_s=3$. We define a grid that covers the FOV with a number of pixels that depends on the resolution of the instrument we are simulating. For example, if we are interested in simulating observations by the HST, we can choose a similar FOV to those of the ACS or WFC3, onboard the orbiting space telescope. The number of pixels follows from the desired resolution, which would be $0.05\arcsec$ in the case of the ACS. In general, the user can adjust both the size of the FOV and the spatial resolution. 

We sum the contributions of all the substructures to determine the complete deflection field. This operation is computationally very expensive, so we perform it adopting the MPI paradigm, following a procedure similar to that presented in the previous section. These maps, which represent the deflections of the cluster for a background source at $z_s = 3$, are rescaled at simulation time by \texttt{SkyLens} to account for sources at different redshifts that are present in the background field. Computing the deflection maps in this framework allows for an improvement in the computation time of more than 10 times compared to the previous one of 60s. Some additional advantages may come from running the simulations on Graphics Processing Units (GPUs). While this possibility is still to be investigated, some GPU-enabled simulators, like \texttt{Caustics} \citep{stone_2024} and \texttt{JAXtronomy} (Huang et al. in prep.) are available in the literature. 

\subsection{Cluster members}\label{sec:clu_memb}
We define the properties of the cluster members according to empirical relations, linking the dynamical properties of the galaxies to the photometric ones, as observed in real clusters. Given the definition of the mass of the halo and redshift, the procedure we implement to generate the properties of the population of cluster members is the following:

\begin{itemize}
    \item We determine the number of galaxies in the cluster by integrating the luminosity function presented in \cite{moretti_2015} up to $R_{200}$ of the cluster. Thus, a more massive cluster will have a larger number of cluster members. To account for some scatter in this value, we assume Poisson fluctuations in the number counts. \cite{moretti_2015} constrain the composite luminosity function of the clusters in the Wide-Field Nearby Galaxy-clusters survey \citep[WINGS;] []{fasano_2006}, observed in the V band. They model the luminosity function as a double Schechter function, written as 
    \begin{equation}
        \phi(L)=\phi^*\Bigg[\Bigg(\frac{L}{L_b^*}\Bigg)^{\alpha_b}\exp\Bigg(-\frac{L}{L_b^*}\Bigg) + \Bigg(\frac{L_b^*}{L_f^*}\Bigg)\times\Bigg(\frac{L}{L_f^*}\Bigg)^{\alpha_f}\exp\Bigg(-\frac{L}{L_f^*}\Bigg)\Bigg].
            \label{eq:lum_func}
    \end{equation}
   Here $L$ is the galaxy luminosity, $L_b^*$, $L_f^*$, $\alpha_b$, and $\alpha_f$ are the characteristic luminosities and slopes of the bright and faint parts of the luminosity function, respectively. This function well characterises both the faint and the bright ends of the luminosity function of the cluster galaxies \citep[e.g.,][]{popesso_2006, lan_2016}. In particular, we use the parameters of best fit in Fig. 8 of \cite{moretti_2015} as inputs of Eq. \ref{eq:lum_func}. This procedure does not account for any dependence of the number of cluster members with the redshift of the cluster, which could slightly affect cluster richness and the small-scale perturbations of the lensing potential, but we might improve on this in future developments of the simulations.

   \item We generate the magnitudes of the cluster members following the luminosity functions presented in \cite{mercurio_2016}. They model the luminosity function of the galaxy cluster MACS J1206.2-0847, combining HST and Subaru observations in the $R_C$ photometric band, as a Schechter function. In particular, they compute different best-fit values of the function for galaxies at different distances from the cluster centre. This is relevant for our simulations because we are interested in the very central regions of the clusters, where the strong lensing events are observed, and using a luminosity function integrated over large radii would be less accurate. In particular, we use the values of the free parameters given in Fig. 2 of their work.

   \item The morphological type of the cluster members (i.e., elliptical, spiral or lenticular galaxies) depends on the redshift and mass of the clusters. \cite{radovich_2020} studied the main properties of the cluster members in the galaxy clusters observed in the third data release of the Kilo Degree Survey \citep[KiDS;][]{kids_2015}. In their work, they schematically divide cluster members into red (elliptical) and blue (spiral) galaxies. They provide the overall fraction of the two morphological types and their radial distributions for clusters of different masses and at different redshifts. We use the distributions shown in Fig. 8 of \cite{radovich_2020} to determine the fraction of elliptical and spiral galaxies in each cluster and Fig. 7 of the same work to model the radial distribution of the galaxies within the cluster. They study the distribution of the cluster members to large distances from the centre (up to 3 times $R_{200}$), but for our work, we only consider the distances within the strong lensing region of the clusters.

   \item Given the morphology of the galaxies, we assign to each one of them a spectral energy distribution (SED). For this step, we use the templates available in the library of the Bayesian Photometric Redshift \citep[BPZ;][]{benitez_2000, benitez_2004} code. These SEDs are used to derive photometric redshifts in \cite{coe_2006}. Several templates are available for the two morphological types we are considering. Although templates are available for starburst galaxies, we do not use them in our simulations, as they are not usually observed in the central regions of galaxy clusters. In the future, we might improve on this aspect by including these templates as well, as it has been shown in some recent works \citep{lagattuta_2022} that starburst galaxies exist over the central few kpc from the galaxy clusters, even though they are not the dominant type ($\sim 10 \%$ of the total population).
   
   \item We model the surface brightness profiles of the cluster members as Sérsic profiles \citep{sersic_1963}, with the corresponding index equal to one for spiral galaxies and four for elliptical galaxies. We assume that the effective radius of this profile and that of the DM profile are related as $r_{eff} = \frac{3}{4} r_{cut}$, where $r_{cut}$ is the characteristic radius of the PIEMD model (eq. \ref{eq:piemd}).

   \item After we have defined the photometric properties of the cluster members, we link them to the dynamical properties, since we need to model their mass distributions, as described in Sect. \ref{sec:mass_model}. For every cluster member $i$ we compute the central velocity dispersion $\sigma_{0, i}$ and cut radius $r_{cut, i}$ following the empirical relations:
    \begin{equation}
        \sigma_{0, i} = \sigma_{0}^{ref}\Bigg(\frac{L_i}{L_0}\Bigg)^\alpha,
        \label{eq:sigma_lum}
    \end{equation}
    and
    \begin{equation}
        r_{cut, i} = r_{cut}^{ref}\Bigg(\frac{\sigma_{0, i}}{\sigma_{0}^{ref}}\Bigg)^\beta,
        \label{eq:radius_sigma}
    \end{equation}
    with the best-fit values of the free parameters given in Fig. 2 and Eq. 9 of \cite{bergamini_2019}.

    \item For the ellipticity of the galaxies, we follow the work by \cite{deugenio_2015}. In particular, we model it according to the distribution of projected ellipticity that the authors found for their reference sample (Fig. 3 of their work.)
    
\end{itemize}

We follow a separate procedure to generate the properties of the  brightest cluster galaxy (BCG), because it has been shown to have different properties with respect to the other cluster members \citep{collins_1998, vonderlinden_2007}. Specifically, we derive the effective radius of the luminous matter of the BCG from the mass of the halo, following the relation
\begin{equation}
    \log (R^{BCG}_{kpc}) = 0.37(\pm0.056)\times\log(M_{cluster}) -3.67(\pm0.81), 
    \label{eq:bcg_radius}
\end{equation}
that was found in \cite{fasano_2010}. We define the other properties of the BCG, namely the magnitude and Sérsic index, following the catalogue in Table 1 of the same work.  We set its PA equal to that of the main cluster's halo. Although some works \citep{roche_2024} have highlighted that the BCG can be offset from the cluster center, particularly in unrelaxed systems, we place it at the center of the halo’s potential well. This is a reasonable assumption for the largely relaxed systems we model and simplifies the initial analysis. Furthermore, we simulate the intracluster light (ICL) as an additional luminous component modeled with a second de Vaucouleurs' profile. \citep[i.e., a Sérsic profile with index equal to one,][]{devaucouleurs_1948} that we use to define the BCG. This approach, that follows the findings of \cite{gonzalez_2005}, might be simplistic, but we find it a good compromise given that we want to simulate a large number of galaxy clusters and we want to avoid to define an ad hoc procedure for every single one of them. An alternative approach would be to model it following the cluster’s gravitational potential, namely using the NFW profile of the cluster's DM halo, given that these stars are expected to broadly trace the potential well of the cluster \citep{montes_2019}. The characteristic radius and luminosity of the ICL are related to those of the BCG \citep{gonzalez_2005}. We assign to the ICL the SED of an elliptical galaxy, assuming that most of its stars are due to the removal of the stars from the cluster members' population \citep[e.g.;][]{morishita_2017} and consists of old stars. While this assumption may be too simplistic in the case of merging clusters, where high ICL fractions and significant contributions in bluer filters indicate the presence of younger or lower-metallicity stars \citep{jimenez_2019}, this approach is sufficiently reasonable for the relaxed clusters considered in our simulations.

In Appendix \ref{app:cl_membs}, Table \ref{tab:clumemb_props} provides a summary of the procedure for generating the cluster members, BCG and ICL in the simulations, as well as the references to the empirical relations we use.

\subsection{Background galaxies}\label{sec:bkg_sources}
We model the background field following the luminosity and redshift distributions of the galaxies in the Ultra Deep Field \citep[HUDF;][]{beckwith_2006}, as given in the catalogue of galaxies presented in \cite{rafelski_2015}. The HUDF covers an area of 10.8 arcmin$^2$ to the limiting magnitude $mag_{AB}\sim29$ for point sources. The images of these galaxies have high spatial resolution and are sufficiently deep to characterise the population of high-redshift sources typically lensed by galaxy clusters. We apply a magnitude cut of 28 on the original catalogue to select the sources for our simulations, resulting in a sample of approximately 2000 objects. 

The cutouts of galaxies we use as background sources were denoised using the expectation-maximisation principal components analysis (EMPCA), and they were presented in \cite{maturi_2017}. This technique uses a linear model to describe the postage stamp of each galaxy $d(\textbf{x})$:
\begin{equation}
    g(\textbf{x}) = \sum_{k=1}^M a_k \phi_k (\textbf{x}) \;\;\;\;\;\;\text{with} \;\;\;\;\;\; a_k = \sum_{1}^n d(\textbf{x}_i) \phi_k(\textbf{x}_i).
    \label{eq:denoising}
\end{equation}
Here, $\{\phi_k \in \mathbb{R} | k =1,.., M\}$ is a set of orthonormal bases that is derived from the data itself, $g(\textbf{x})$ is the de-noised model of the galaxy, and $n$ is the number of components, that is equal to the number of pixels of the image. Two components make up the data:
\begin{equation}
    d(\textbf{x}) = g(\textbf{x}) + n(\textbf{x}),
    \label{eq:image_noise}
\end{equation}
where $n(\textbf{x})$ is the model of the noise. The EMPCA uses the information content of the images to define the $M$ components with more information than the others to describe the image. Thus, it produces a de-noised image of the original postage stamp for each galaxy. More details about this method are in \cite{maturi_2017}.

The SEDs of the galaxies in the HUDF have been studied in \cite{coe_2006}: we assign to each of the sources we select the best-fit SED template from their work, re-scaling it at the redshift of the sources. Finally, we randomly position the postage stamps of the background galaxies in the FOV of our simulated galaxy cluster, rotating and flipping them to obtain a different realisation of the HUDF for every image we generate.

In addition to the real observations, we use 1000 additional images generated with a score-based diffusion model, to prevent the same galaxies from appearing multiple times in the FOV and ensure variation in the simulations. Score-based diffusion models are a class of Machine Learning (ML)-based generative models that were introduced in \cite{song_2021}, and became increasingly popular in recent years for their effectiveness in learning the underlying distribution of the input dataset with no prior knowledge of it. These models can generate new samples through de-noising, namely, they are trained by adding random noise to the input data and learn to progressively de-noise it to re-obtain the target data distribution. 

In particular, at training time, the noise is added to real data in many incremental steps, hence producing a series of increasingly noisier versions of the input. For each noise level, the model learns a score function, that is, the gradient of the probability distribution function of the input data. At each step, the score function essentially indicates how to de-noise the noisy versions of the input data to have a new (less noisy) version closer to the true input distribution. At generation time, this process is reversed: the model starts from pure random noise and gradually removes it following the learned score function, thus producing images that capture the complexity of the data.

For our case, we apply conditional training to guide the model in generating galaxies with the specific characteristics we are interested in, i.e. the redsfhit and color of the galaxies. For this reason, we adopt the framework presented in \cite{adam_2022} for training the original score-based diffusion model of \cite{song_2021} on the images of the HUDF, conditioning on the redshift and SED types of the galaxies. By incorporating these additional requirements into the training procedure, the model learns to associate specific noise patterns with some conditions (e.g., galaxies at particular redshifts or with a given SED type). The training setup and code are publicly available on GitHub at this \href{https://github.com/AlexandreAdam/hst_diffusion_project}{link}. 

When generating new samples, we input the SED type and redshift into the model, enabling it to generate images of galaxies that are realistic and also follow the redshift and SED characteristics of the HUDF. This ensures that the generated galaxies reflect the distribution of the original data but allows for more variation of the images in the background field.

\begin{figure*}
    \centering
    \includegraphics[width = 0.95\textwidth]{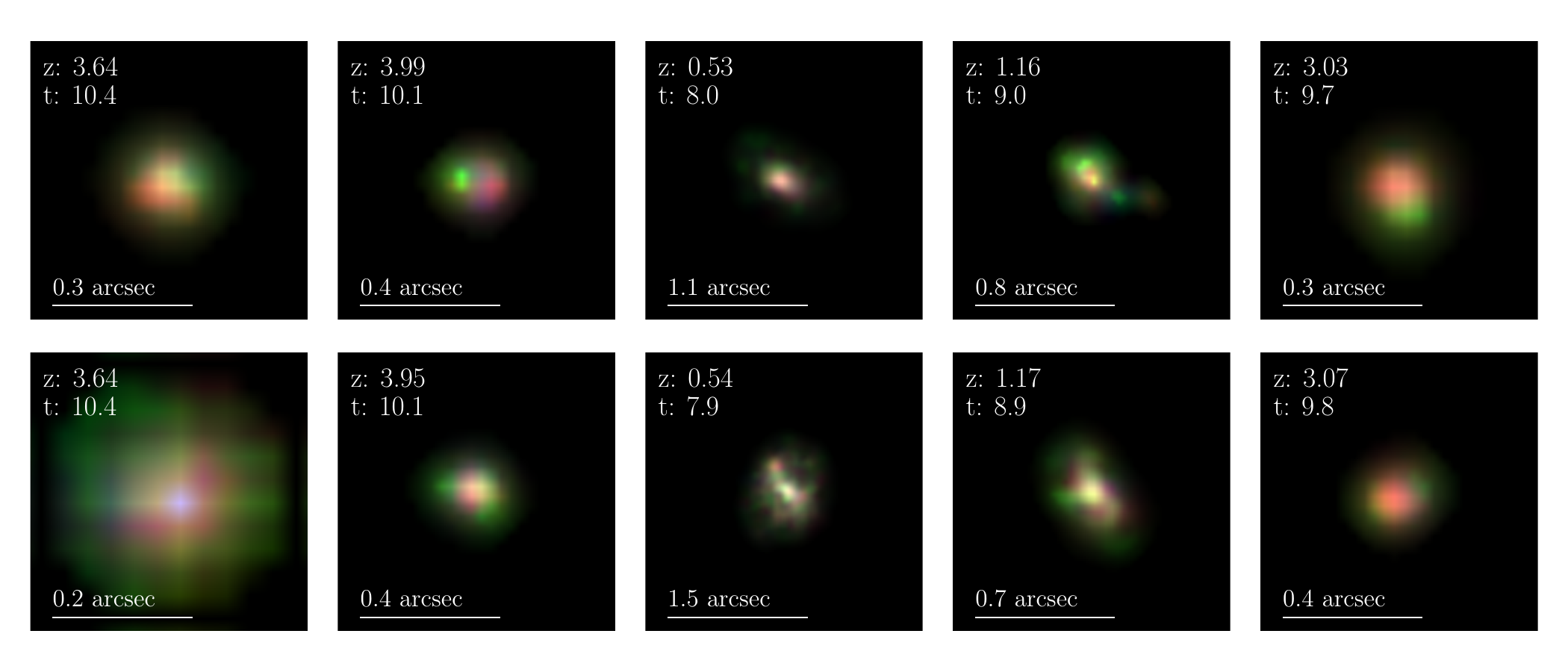}
    \caption{In the upper row, we show five galaxies observed in the HUDF, while in the second row, we present five examples of the images generated by our score-based diffusion model. In each panel of the Figure, we report the redshift and SED template of the galaxy. These are two properties on which the diffusion model was conditionally trained.}
    \label{fig:bkg_sources}
\end{figure*}

\begin{figure*}
    \sidecaption
    \includegraphics[width=12cm]{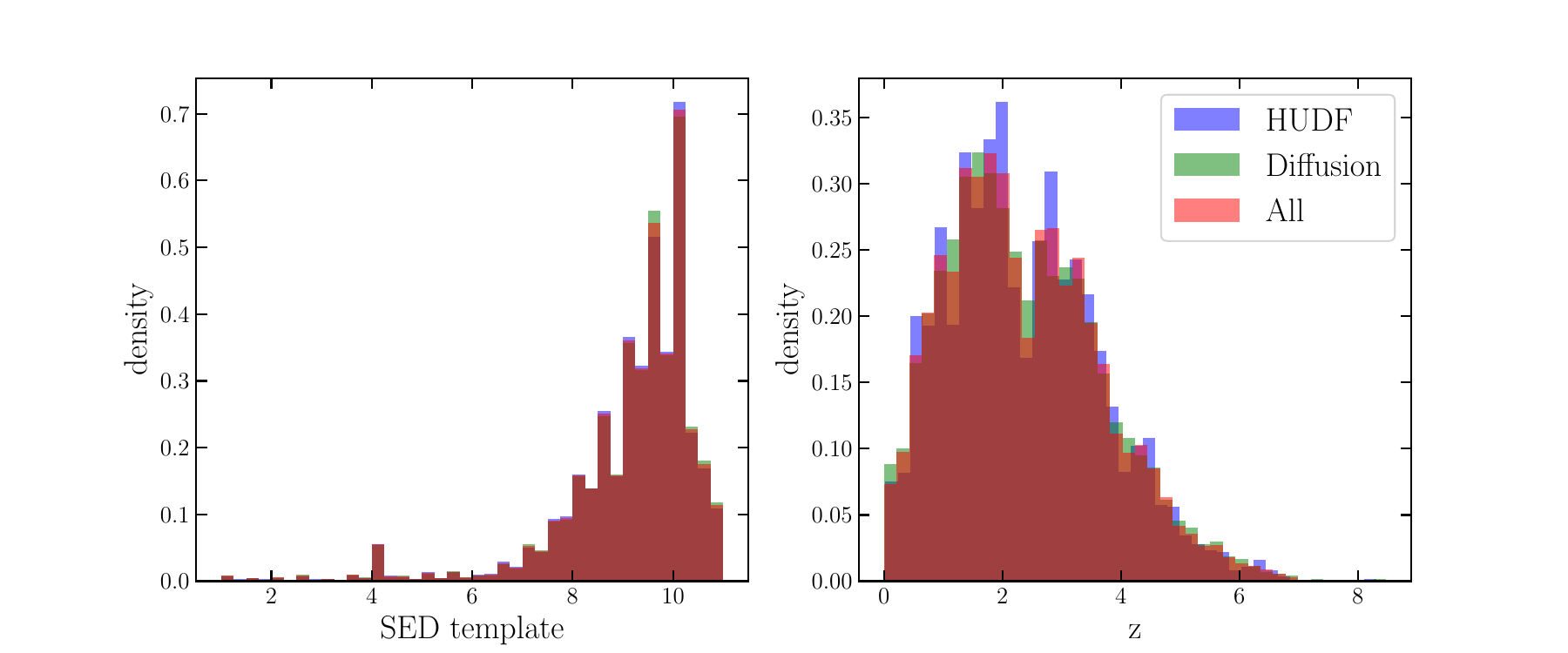}
    \caption{Comparison of the spectral type (left panel) and redshift (right panel) distributions for the real and simulated galaxies. Since we conditionally trained the diffusion model on these two properties, the two distributions are very similar. In the first panel, the spectral type is defined as a number in the range 0 to 11: this corresponds to the possible templates in BPZ \citep{benitez_2000, benitez_2004}. In particular, the numbers close to zero refer to elliptical galaxies, while those close to eleven describe spiral galaxies.}
    \label{fig:bkg_hists}
\end{figure*}

In Fig. \ref{fig:bkg_sources} we show five original de-noised with EMPCA postage stamps from the HUDF (first row) and five mocks generated by our diffusion model (second row). In each panel, we show the redshift and SED numbers of the galaxy. Moreover, in Fig. \ref{fig:bkg_hists} we show the distribution of the SED types and redshift of the sources in the original data set and in the mock one.

The catalogue of properties of the background sources includes different entries from those of the cluster members: we include the redshift of the sources, their magnitude in each band of the simulation, information about the flipping and rotation, and the position in the FOV. In particular, in the case of the lensed background sources, we compute the positions of all the multiple images and save them in the catalogue as well.

\subsection{Mock observations}\label{sec:mock_obs}
As explained, the images simulated by \texttt{SkyLens} are convolved with the PSF of the specified instrument, but they are noiseless. In the final step, we add noise to the images using some proper functions from the \texttt{pyLensLib} package. Given the main characteristics of the instrument and of the observational setup we are simulating (zero point, FOV, exposure time), this package computes the background level of the sky and the noise of the image, either using Gaussian or Poisson noise (depending on the choice specified by the user).

\begin{figure*}
    \sidecaption
    \includegraphics[width=12cm]{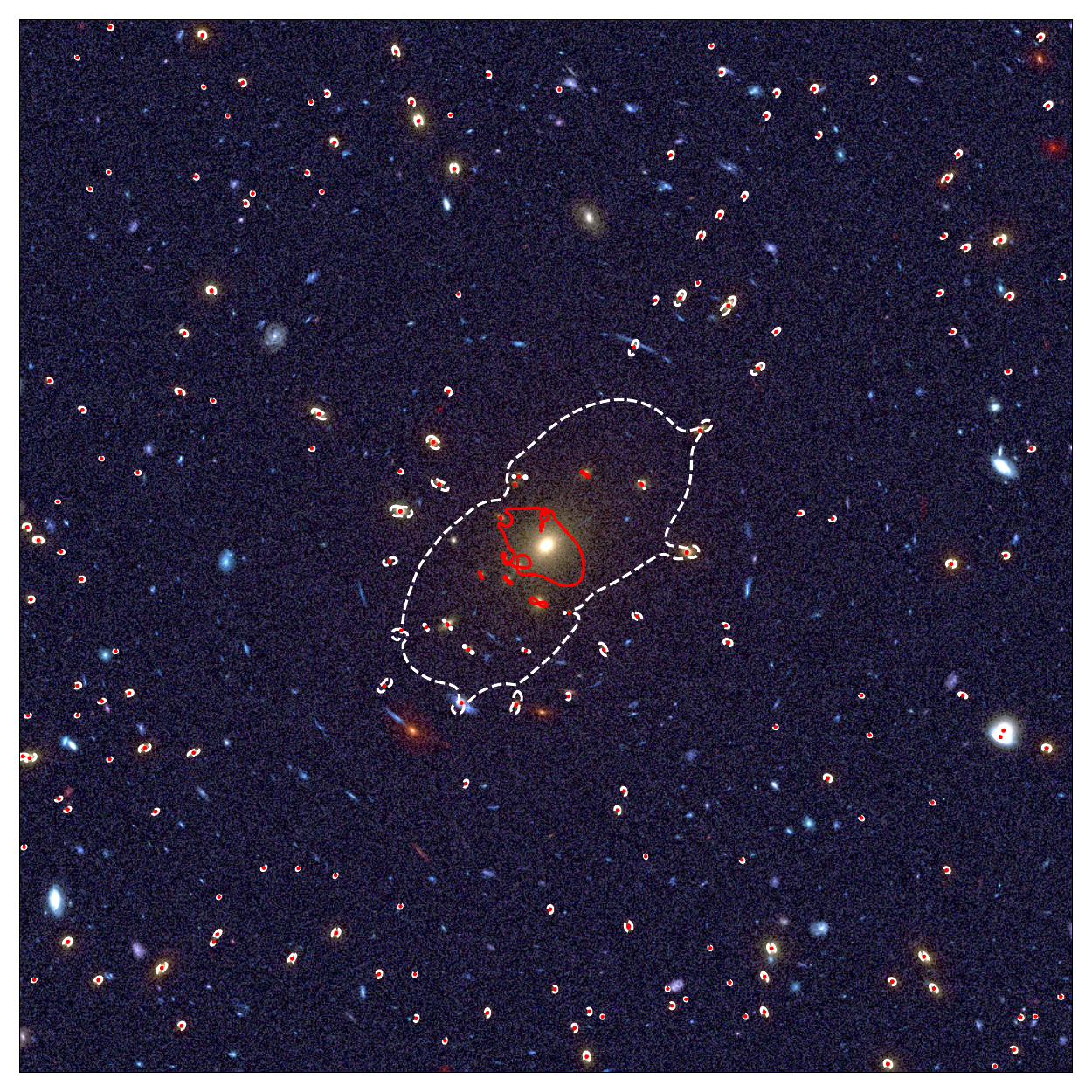}
    \caption{Example of a simulated galaxy cluster. This is a composite image of the F850LP, F606W, F475W bands of the HST. We superimpose on the image the critical lines of the main dark matter halo and those of the cluster members. In white dashed lines are the tangential critical lines, while in red solid lines are the radial critical lines. This simulation covers a FOV of $200\arcsec\times200\arcsec$, with a spatial resolution of $0.05\arcsec$.}
    \label{fig:criticallines}
\end{figure*}

\begin{figure}[t]
    \centering
    \includegraphics[width = 0.49\textwidth]{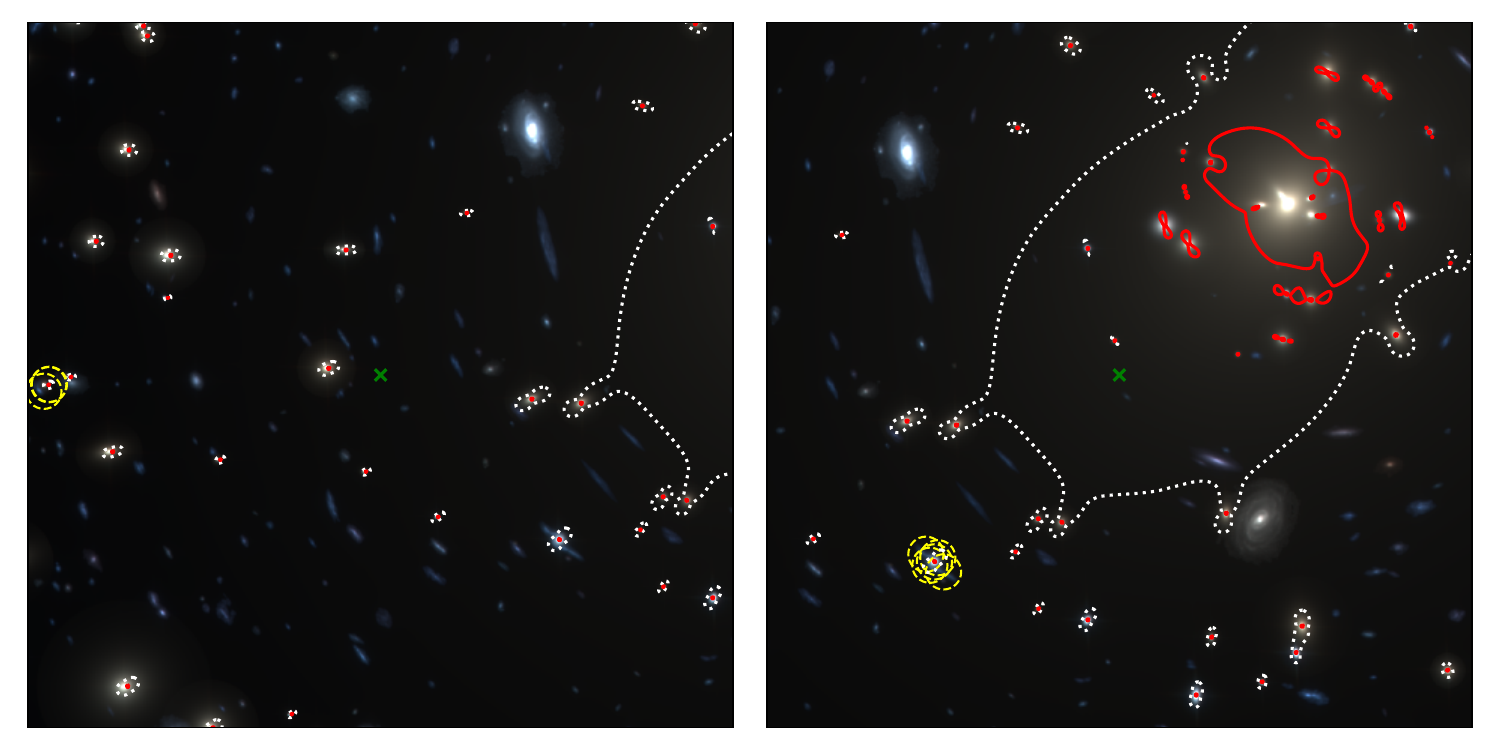}    
    \caption{Two examples of configurations of multiple images of a background source in a HST-like simulation. In each panel, we show a cutout of the area where the multiple images (circled in the dashed yellow lines) are. The cutouts are centered on the unlensed position of the background source, that we mark with the green cross.} We superimposed on the image the tangential (dotted white line) and radial (solid red line) critical lines of the system, which mark the cluster regions that would infinitely magnify a background source.
    \label{fig:multiple_img}
\end{figure}

We show in Fig. \ref{fig:criticallines} an example of a simulated cluster of mass $M_{200} = 1.5 \times 10^{15} M_\odot$, at $z = 0.4$. We superimpose the primary and secondary critical lines of mass distribution on the image. This simulation covers a FOV of $200\arcsec\times200\arcsec$, with a spatial resolution of $0.05\arcsec$ (hence, it is a HST-like simulated image). Furthermore, we show in Fig. \ref{fig:multiple_img} the cutouts of two cluster regions in a different simulated cluster, in which multiple images of lensed background sources (circled in the yellow dashed lines) appear. We show some additional examples of simulated clusters in Appendix \ref{app:examples}.

\section{Validation}\label{sec:frontierfields}
The internal functions implemented in \texttt{SkyLens} for the ray-tracing, the distortion of the background sources, the multiplane-lens configuration, have been validated in several works in the literature \citep[see][and references therein]{skylens_plazas}. Here, we validate our choices for generating the cluster members, creating the mass model, and simulating the background sources. We do so by simulating a galaxy cluster with properties similar to those of MACS J1206.2-0847 (hereafter, MACS1206) and comparing the properties of the cluster members and lensing signals. 

MACS1206 is a known and well-studied cluster-lens at redshift $z = 0.44$. It was observed as part of the CLASH program, which targeted 25 massive galaxy clusters with the objective of mapping their dark matter distribution, studying their internal structure, and using their magnifying power to detect faint and distant sources. Thanks to these observations, several properties of MACS1206 have been investigated in detail, including the properties of its member galaxies.  For example, \cite{annunziatella_2016, kuchner_2017, ferrami_2023} have characterised the population of cluster members, derived their stellar mass function, and studied the effect of the environment on their morphology. Moreover, \cite{biviano_2023} have constrained the distribution of dark matter in the internal regions of the cluster using galaxy dynamics. Several works \citep[e.g.,][]{zitrin_2012, umetsu_2012, caminha_2017} have addressed the study of the internal mass distribution of the cluster by leveraging strong and weak lensing measurements. All the available observations and mass models of the clusters targeted in the CLASH program, including MACS1206, are publicly available at this \href{https://archive.stsci.edu/prepds/clash/}{link}.

In our work, we consider the two main diffuse mass component distributions of the lens model presented in \cite{caminha_2017} as the input of our simulation. This is a simplified assumption as we neglect one additional diffuse mass component and the external shear contribution. We neglect this third diffuse component as it has a velocity dispersion 30\% lower than the main component, making it negligible. Whereas the external shear component is neglected because it only marginally affects the strong lensing features we are aiming at, i.e. the position and shape of the multiple images observed in cluster cores. In these regions, the effect of the shear is subdominant, thus we have neglected it. Moreover, our purpose is not to replicate the same exact mass distribution of MACSJ1206 but to verify that, assuming the same redshift and total mass, our code produces a mock galaxy cluster with comparable galaxy population and lensing signal (i.e., with a similar Einstein radius and lensing cross section), to the real one. 
We present a summary of the parameters that we use to define the two considered components in Table \ref{tab:m1206}, which is based on Table A.2 of \cite{caminha_2017}. In particular, we use the coordinates of the centres of the main haloes ($x,y$), their axis ratio $q$ and the central velocity dispersion $\sigma_0$, as a proxy of the halo's mass. We make no assumptions about the cluster members' population, but use the procedure described in Sect. \ref{sec:clu_memb} to generate their properties.

\begin{table}
\centering
\caption{Summary of the parameters of the two main diffuse mass components used to simulate the MACS1206-like cluster.}
\begin{tabular}{ccc}
&Main component& Second component\\ \hline
$x\, (\arcsec)$&-0.5& 8.6\\ 
$y\,(\arcsec)$&0.4&4.4\\
$q$&0.44&0.62\\
$\sigma_0\,(\text{km\,s}^{-1}$)&951&863\\
\hline\\ [-1.8ex]
\end{tabular}
\tablefoot{The values in this Table are based on those in Table A.2 of \cite{caminha_2017}. We report here the coordinates ($x$ and $y$, in arcsec) of the centres of the two haloes with respect to the centre of the image, the axis ratio $q$, and the central velocity dispersion $\sigma_0$.}
\label{tab:m1206}
\end{table}

\begin{figure}
    \centering
    \includegraphics[width=0.99\linewidth]{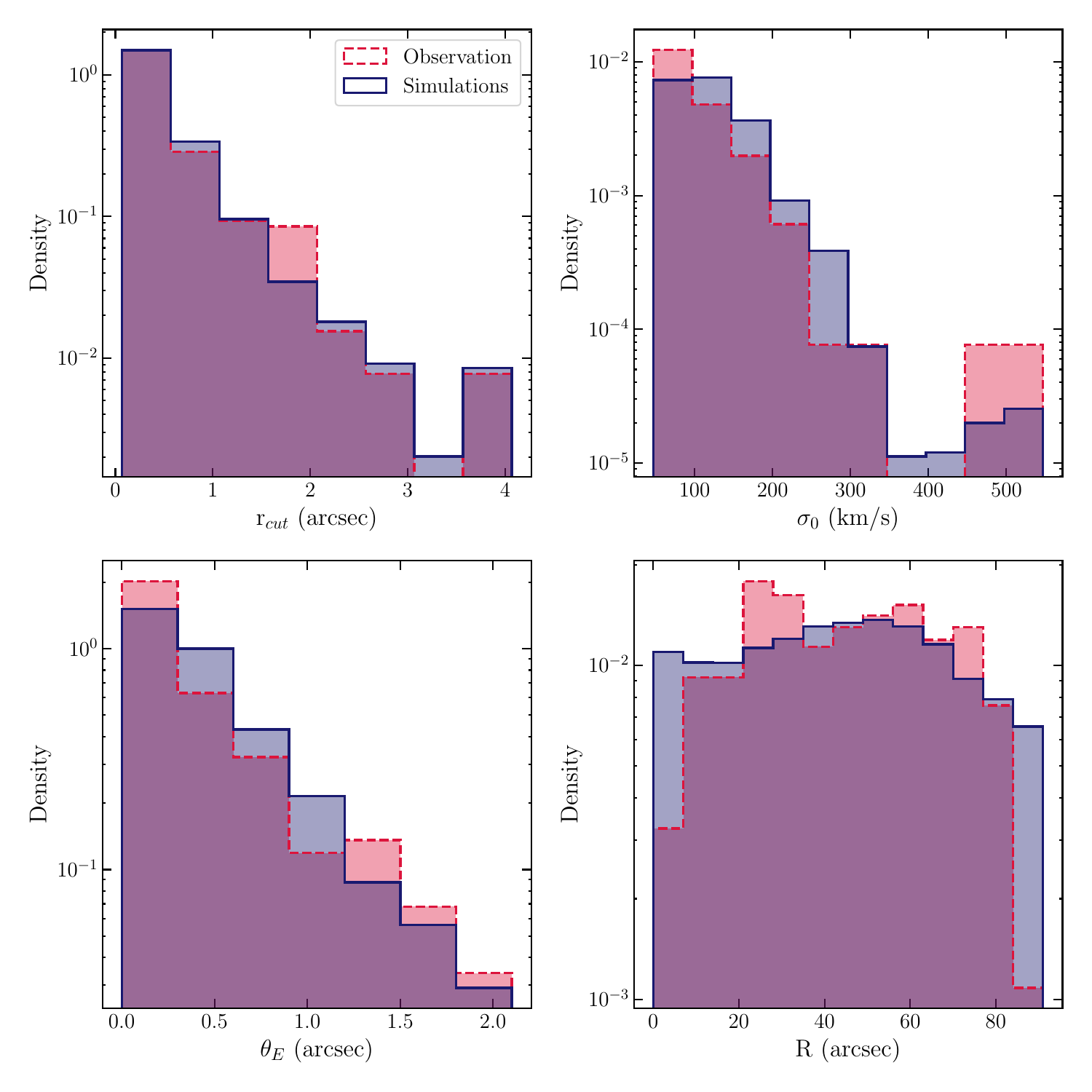}
    \caption{Comparison of relevant properties of the cluster members observed in MACS1206 (in red, dashed line) and in our simulation of the same cluster (in blue, solid line). From the top-left to the bottom-right panel, we show the distributions of galaxy size ($r_{cut}$), velocity dispersion ($\sigma$), Einstein radius ($\theta_E$), and radial position. For the radial distribution, $R$ denotes the distance from the cluster centre. These histograms are based on the average of the realisations of our simulation.}
    \label{fig:comparison_clmembs}
\end{figure}

We can compare the results of our simulations with the observations by taking into account some key quantities.
\begin{enumerate}
    \item In \cite{caminha_2017}, the cluster members have been selected with the limiting magnitude of 24 in the F160W band of the HST. Following the same criterion, we integrate the luminosity function of \cite{moretti_2015} down to this magnitude cut, as explained in the previous Section. The number of simulated cluster members slightly varies in different runs, with an average value of $N_{cl} = 240$, to which we add some dispersion according to a Poissonian distribution, thus introducing a scatter of $\sim \pm 15$. This number should be compared to 264 cluster members used in the mass model by \cite{caminha_2017}.

    \item The main critical line of our simulated cluster, as described in Sect. \ref{sec:clusters_props}, eq. \ref{eq:ein_radius}, has Einstein radius $R_E = 25\arcsec$, while that of MACS1206 has $R_E = 31\arcsec$. Thus, our simulation predicts an Einstein radius that is $\sim20\%$ smaller than the observed one. Given that the purpose of the simulation is not to reproduce the detailed mass distribution of MACS1206, we consider this level of agreement satisfactory for the scope of this work.
    
    \item In Fig. \ref{fig:comparison_clmembs} we show the comparison of the population of cluster members in our simulated MACS1206-like cluster and in the real cluster. In the first row, we compare the radius of the galaxies, hence their size, and their velocity dispersion, which is a proxy of their mass. In the second row, we compare the Einstein radii distribution of the secondary critical lines of the cluster members and their radial distribution function. In all these quantities, we find good agreement between the observations and the simulations. These histograms show the results of the average of the realisations of our simulation.
\end{enumerate}

\section{Discussion}\label{sec:discussion}
The main application of the simulation code presented in this work is to generate training sets of images and catalogues for ML algorithms, for which a large number of labelled images is needed. Although the application of ML methods in recent years has mostly been tested for the identification \citep[e.g.,][]{jacobs_2019, canameras_2021, rojas_2022, savary_2022} and modelling \citep[e.g.,][]{hezaveh_2017, gentile_2023, biggio_2023} of galaxy-scale strong lenses, characterising the applicability of the same methods to lensing by galaxy clusters is equally essential. 

Investigating strong lensing by galaxy clusters is very time-consuming because of their physical complexity compared to galactic lenses and the poor automation of the techniques to date. In this context, some groups have started to investigate the applicability of DL (Deep Learning) methods to some relevant tasks. \cite{jia_2022} apply transformers to identify lensed arcs in galaxy clusters, using the simulations produced with the code PICS \citep{pics_simulations}. Their approach shows promising results on simulations ($90.32\%$ recall and $85.37\%$ precision, with a false positive rate of just $0.23\%$), but it has not been applied to real data yet. \cite{angora_2020} have investigated the use of deep CNNs to identify cluster members in real observations of galaxy clusters by the HST. More recently, \cite{angora_2023} have also applied DL to identify galaxy-scale lenses in clusters. Both these methods might be relevant when modelling strong lensing by galaxy clusters. Nonetheless, the path to having a reliable and automated pipeline in this field is still long.  

The use of simulations is necessary not only for training ML models but also for validating modelling and image-reconstruction codes and characterising their biases and systematic errors in retrieving lens parameters. Given that the exact shape of the lensed background sources is not known a priori, the use of simulations with known input is an important validation step \citep[e.g.,][]{tessore_2016, denzel_2020, ding_2021} before applying novel  reconstruction methods to real observations. For example, \cite{meneghetti_2017} used two clusters simulated with \texttt{SkyLens} in different dynamical states to compare the results of independently developed mass reconstruction methods.

Comparing simulations and observations is also essential to determine the accuracy of the assumptions on which the simulations are based \citep[e.g.,][]{mazzotta_2004, pratt_2010, borgani_2011, gatuzz_2025}, and potentially rule out some cosmological scenarios over others. This is the case of the work by \cite{meneghetti_2020} \citep[and the subsequent ones by][]{ragagnin_2022, meneghetti_2022, meneghetti_2023}, that have highlighted a significant discrepancy in the number of observed and expected galaxy-scale strong lenses in galaxy clusters. The number of these events in the simulations depends on the assumed dark matter model, and the comparison between the predictions of different paradigms and observations might be decisive in favouring one scenario over another. \cite{mastromarino_2023} also investigated alternative dark matter models and their impact on the Einstein radii distribution of lenses. 

All the applications mentioned so far require highly realistic simulations. In the future, we plan on improving the realism of our simulations by introducing some refinements and extensions to our methodology. One possible improvement is the representation of the background sources. As discussed in Sect. \ref{sec:bkg_sources}, we have relied on the galaxies observed in the HUDF, but we plan on testing the difference we would have when using observations of the JWST. In particular, this would improve the representation of background sources in the redder bands and would enable the inclusion of higher-redshift galaxies, thanks to the greater sensitivity and longer-wavelength coverage of JWST compared to HST. In the same Section, we have also explained some of the advantages of score-based diffusion models for augmenting the background sources. In addition to generating new images, this technique is also promising in simulating realistic noise distributions \cite[e.g.,][]{legin_2023}. We plan on applying it to these stages of the simulations to make them faster and more realistic. 

In the future, we also plan on releasing larger sets of simulations and corresponding catalogues for the community. In this regard, it would be important to consider clusters in a perturbed dynamical state, i.e., merging clusters, as well as relaxed clusters. In this case, the deflection maps would have to include multiple dark matter haloes and be modified to account for the displacement of the cluster members. Moreover, we will include second-order perturbations to the deflection fields of the clusters to account for external structures that might impact the way the path of the photons of the background sources is deflected. 

\section{Conclusions}\label{sec:conclusions}
In this work, we have presented a new procedure to generate cluster-scale strong lenses, which is based on \texttt{pyLensLib} \citep{meneghetti_libro} and \texttt{SkyLens} \citep{skylens_plazas}.  With our procedure, we have simulated the images and the catalogues of several properties of the clusters (see Sect. \ref{sec:clusters_props}), cluster members (see Sect. \ref{sec:clu_memb}), and background sources(see Sect. \ref{sec:bkg_sources}). In particular, we release with this paper a set of one hundred simulations, along with the catalogues of several properties of the clusters, cluster members and background sources. These catalogues are useful to train ML and DL methods to identify and characterise different properties of the clusters because they have information about the lensing, photometric, and dynamical properties of every source.  

We have simulated one hundred clusters, and we have outlined their properties. They are simulated in a mass range of $M_{200} \sim 5\times10^{14} - 2\times10^{15}M_{\odot}$, and are located in the redshift range $0.3-0.6$. For this simulated set, we use empirical relations that were derived from real observations to determine the properties of the cluster members, the BCG and the ICL to ensure the realism of the underlying physical properties of the galaxy clusters. We simulate the background field using postage stamps of the galaxies observed in the HUDF that have been de-noised with the EMPCA decomposition method by \cite{maturi_2017}. In addition to the real images, we generate more cutouts with a score-based diffusion model \citep{adam_2022} that we trained on the HUDF observations. Thus, we have different realisations of the HUDF for every image we simulate, ensuring that the background field is not the same for every simulated mock we create.

Our simulations are mostly based on two codes. We generate the deflection angles and the catalogues of cluster members and background sources,  and we simulate the observational effects using \texttt{pyLensLib} \citep{meneghetti_libro}, a Python-based package developed to simulate and analyse lensing events. Moreover, we use \texttt{SkyLens} \citep{skylens_plazas} to perform the ray-tracing and generate the images using the catalogues and deflections as input. The main modifications we introduced in these codes aim to reduce the computation time to simulate thousands of images in a reasonable amount of time (namely, a few days). We did this by implementing the computation of the deflection maps and the ray tracing in an MPI framework, where the grid on which these calculations are made is split into several processes carried out in parallel. Following this paradigm, we have reduced the computation time of our simulations by a factor of ten.

We validate our simulations by comparing the physical properties and lensing signal of a simulated cluster that has similar mass and redshift of MACS 1206, to those derived by the real observations of the cluster. Our simulations are highly relevant for a range of applications, including the training of ML-based algorithms for strong lensing studies, the validation of traditional modelling techniques, and the assessment of the accuracy of simulation assumptions through comparison with observations. We make this data set and corresponding catalogues available to the community at this \href{https://openscience.inaf.it/?page_id=724}{link}.

\begin{acknowledgements}
LL and CG thank the support from the INAF theory Grant 2022: Illuminating Dark Matter using Weak Lensing by Cluster Satellites. LM acknowledges the financial contribution from the 
PRIN-MUR 2022 20227RNLY3 grant “The concordance cosmological model: stress-tests with galaxy clusters” supported by Next Generation EU and from the grant ASI n. 2024-10-HH.0 “Attività scientifiche per la missione Euclid – fase E”. The work of AAPM was supported by the U.S. Department of Energy under contract number DE-AC02-76SF00515. MM was supported by INAF Grants “The Big-Data era of
cluster lensing". PB acknowledges financial support through grant PRIN-MIUR 2020SKSTHZ and support from the Italian Space Agency (ASI) through contract ``Euclid - Phase E'', INAF Grants ``The Big-Data era of cluster lensing'' and ``Probing Dark Matter and Galaxy Formation in Galaxy Clusters through Strong Gravitational Lensing''.

\end{acknowledgements}

\bibliographystyle{aa}
\bibliography{paper}
\begin{appendix}
\onecolumn
\section{Additional examples}\label{app:examples}
\begin{figure}[h!]
    \centering
    \includegraphics[width=0.95\linewidth]{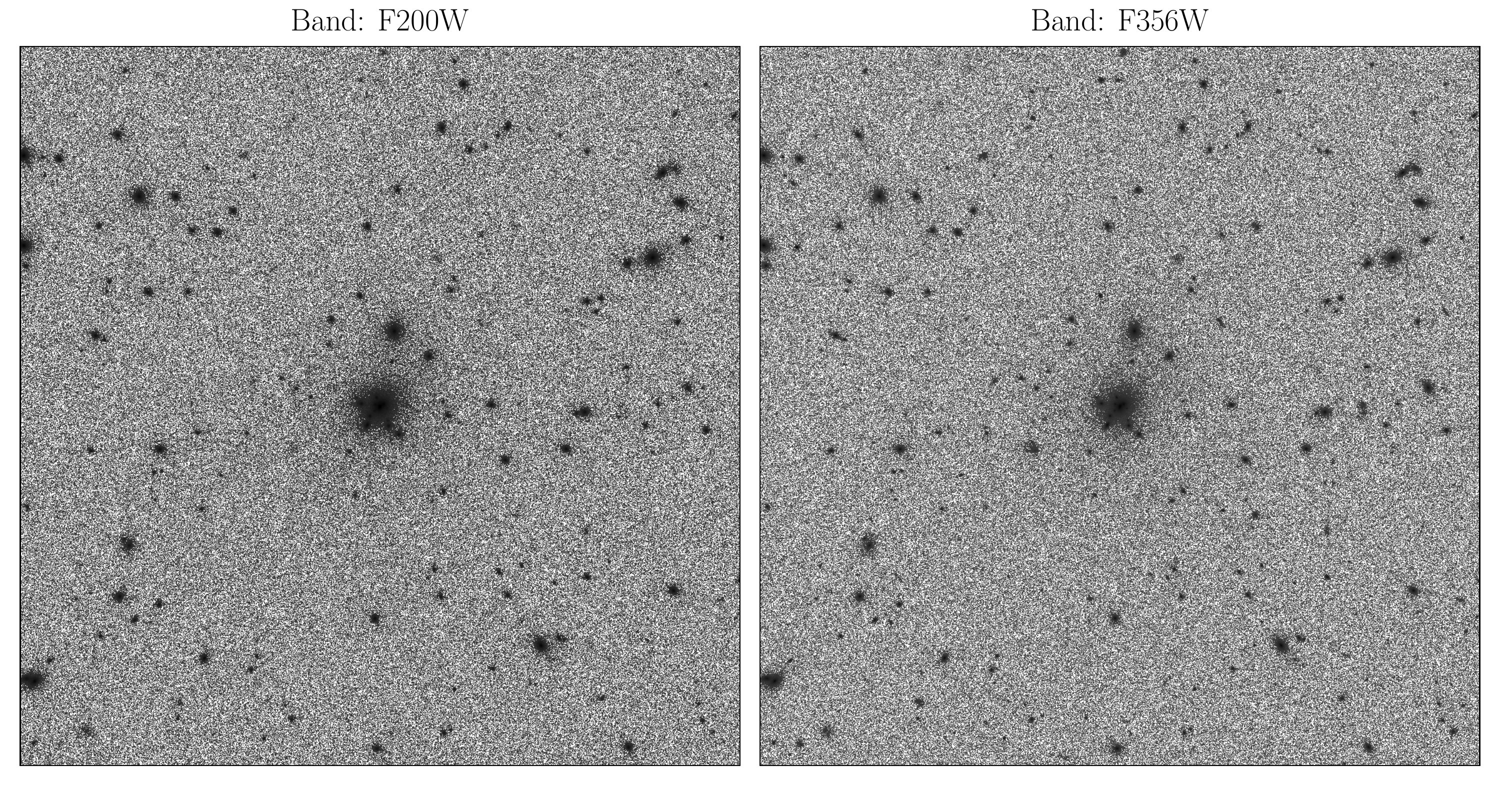}
    \caption{Example of a galaxy cluster simulated in the F200W and F356W bands on the NIRCam instrument of the JWST.}
    \label{fig:jwst_example}
\end{figure}

\begin{figure}[h]
    \centering
    \includegraphics[width=0.95\linewidth]{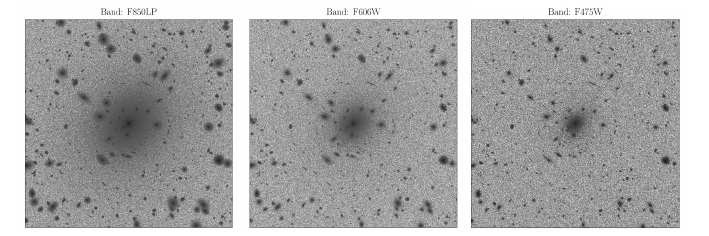}
    \caption{One-band images of the simulated HST-like cluster shown in Fig. \ref{fig:criticallines}. In the first row, we show the cluster as simulated in the F850LP (left) and F606W (right) bands, while in the second row we show it in the F475W band.}
    \label{fig:one-band-examples}
\end{figure}

In this Appendix, we show additional examples of simulated galaxy clusters. In particular, in Fig. \ref{fig:jwst_example} we show a galaxy cluster simulated in the F200W and F356W bands of the NIRCam instrument of the JWST, with the noise parameters given in Table \ref{tab:data_noise}. In Fig. \ref{fig:one-band-examples}, we show the one-band images of the same cluster shown in Fig. \ref{fig:criticallines}.

\clearpage
\section{Summary of cluster members' properties}\label{app:cl_membs}
In Table \ref{tab:clumemb_props} we summarize the empirical relations used to simulate the properties of the cluster members, BCG and ICL and point to the corresponding references.
\begin{table*}[h]
\centering
\caption[Summary of the empirical relations we use for the simulated clusters.]{Summary of the procedure we follow to generate the main properties of the cluster members, BCG and ICL. In the last column, we also give the reference to the works that present the empirical relations we use in the simulations.}
\begin{tabular}{ccc}
&Property& Reference\\ \hline
&& Integration of the luminosity function of \\
&Number&\cite{moretti_2015}. It depends on the size of the \\
&&main halo and limit magnitude of the simulation.\\
&&\\
&&Luminosity functions of \cite{mercurio_2016}.\\
&Magnitude&The luminosity functions are given in bins of\\
&&distance from the centre of the cluster, with\\
&& limiting magnitude 24 in the F160W band of the HST.\\
&&\\
&& Overall number of red (elliptical)\\
&Morphology&and blue (spiral) galaxies from \cite{radovich_2020}.\\
&&They depend on the redshift and mass of the main halo.\\
&&\\
&&Radial distribution functions of \cite{radovich_2020}.\\
Cluster members & Position & They depend on the redshift and mass of the main halo\\
&&and on the morphology of the galaxies.\\
&&\\
&&Templates available in BPZ\\
&SED&\citep{benitez_2000,benitez_2004}\\
&&They depend on the morphology of the galaxies.\\
&&\\
&Surface brightness&Sérsic profiles \citep{sersic_1963}, with\\
&profile&Sérsic index dependent on the morphology\\
&&of the galaxies.\\
&&\\
&Velocity&Relation between the velocity and luminosity\\
&Dispersion&in \cite{bergamini_2019}.\\
&&\\
&Radius of&Relation between the radius and velocity\\
&DM halo&in \cite{bergamini_2019}.\\
&&\\
&Ellipticity&Ellipticity distribution function\\
&&in \cite{deugenio_2015}.\\
\hline\\ [-1.8ex]
&Size&Relation between the radius and mass of the main halo\\
&&in \cite{fasano_2010}.\\
BCG&&\\
&All other&From the catalogue in \cite{fasano_2010}.\\
&properties&They depend on the size of the BCG.\\
\hline\\ [-1.8ex]
ICL&&Second component of BCG's Sérsic profile.\\
&&Relations in \cite{gonzalez_2005}.\\
\hline\\ [-1.8ex]
\end{tabular}
\label{tab:clumemb_props}
\end{table*}
\end{appendix}
\end{document}